\documentclass[twocolumn,prb,showpacs]{revtex4}

\usepackage{graphicx}
\usepackage{bm}
\newcommand{\RbFe}{RbFe(MoO$_4$)$_2$}

\begin{document}
\preprint{APS/123-QED}

\title{Quasi two-dimensional antiferromagnet on a triangular lattice \RbFe}

\author{L.~E.~Svistov, A.~I.~Smirnov, L.~A.~Prozorova}
\affiliation{P.~L.~Kapitza  Institute for  Physical  Problems RAS, 117334 Moscow, Russia}

\author{O.~A.~Petrenko}
\affiliation{Department of Physics, University of Warwick, Coventry, CV4 7AL, UK}

\author{L.~N.~Demianets, A.~Ya.~Shapiro}
\affiliation{A.~V.~Shubnikov  Institute for  Crystallography  RAS, 117333 Moscow, Russia}
\date{9 Oct 02}

\begin{abstract}

\RbFe\ is a rare example of a nearly two-dimensional Heisenberg
antiferromagnet on a triangular lattice.  Magnetic resonance
spectra and magnetization curves reveal that the system has a
layered spin structure with six magnetic sublattices. The
sublattices within a layer are arranged in a triangular manner
with the magnetization vectors 120$^\circ$ apart. The $H-T$ phase
diagram, containing at least five different magnetic phases is
constructed. In zero field, \RbFe\ undergoes a phase transition at
T$_N=3.8$~K into a non-collinear triangular spin structure with
all the spins confined in the basal plane. The application of an
in-plane magnetic field induces a collinear spin state between the
fields $H_{c1}$=47 kOe and $H_{c2}$=71 kOe and produces a
magnetization plateau at one-third of the saturation moment. Both
the ESR and the magnetization measurements also clearly indicate
an additional first-order phase transition in a field of 35 kOe.
The exact nature of this phase transition is uncertain.

\end{abstract}

 \maketitle

\section{Introduction}

The problem of an antiferromagnet on a triangular planar lattice
(AFMT) has been intensively studied theoretically.
\cite{Kawamura,Korshunov,Anderson,PlumerCalie,Chubukov} The ground
state in the Heisenberg and XY-models is a ``triangular" planar
spin structure with the three magnetic sublattices arranged
120$^\circ$ apart. The orientation of the spin plane is not fixed
in the exchange approximation.

A non-zero magnetization appears in the presence of a magnetic
field due to the canting of the sublattices. Possible
field-induced structures are shown on Fig.~1. All configurations
with equal magnetization vectors but with different sublattice
orientations have the same energy in the molecular field
approximation. \cite{Chubukov} The umbrella-like structure ``a"
with the sublattices tilted from the spin plane towards the field,
and the planar structures ``b" and ``b$^\prime$" are among these
degenerate configurations. In this approximation the
b-configuration becomes collinear (c-configuration) at a
particular field $H_c=H_{sat}/3$, where $H_{sat}$ is the
saturation field. In magnetic fields above $H_c$, the structure is
again noncollinear, with two parallel sublattices tilted with
respect to the third, forming the ``canted" d-phase. Finally, at
the saturation field $H=H_{sat}$, a spin-flip transition to the
phase ``f" occurs.

\begin{figure}
\includegraphics[width=\columnwidth]{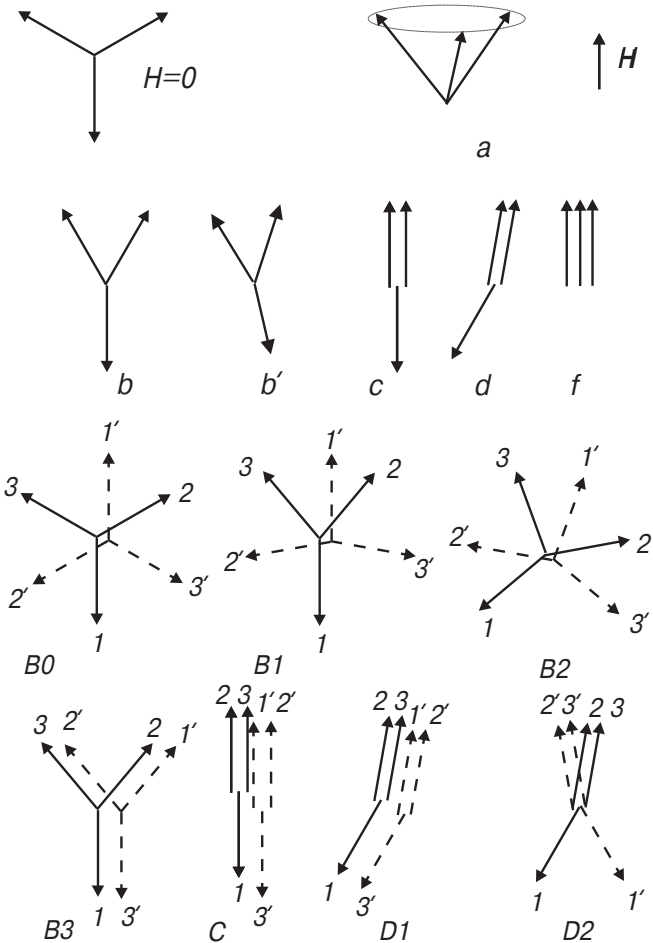}
 \caption{ Schematic representation of the proposed
spin structures of a Heisenberg antiferromagnet on a triangular
lattice. Structures a, b, b$^\prime$, c, d and f are related to
the 2D system ($J^\prime=0$). Structures B0, B1, B2, B3, C and D
represent the triangular antiferromagnet with a weak
antiferromagnetic interlayer exchange in the six-sublattice
model:~\cite{Gekht} solid  and dashed arrows with the same numbers
correspond to magnetic moments of neighboring spins from
neighboring layers.} \label{fig:structure}
\end{figure}

Because of the degeneracy of the classical spin configurations,
both quantum and thermal fluctuations play an important role in
the formation of the equilibrium state of the AFMT.
\cite{Chubukov,Rastelli12,Korshunov} They result in a free-energy
gain of the planar structure with respect to the umbrella-like
structure. Due to the contribution of fluctuations to the free
energy the more symmetric configuration ``b" is preferred to
``b$^\prime$". Further, the fluctuations stabilize the collinear
spin configuration ``c" in a range of magnetic fields $H_{c1} \leq
H \leq H_{c2}$ around the special point $H=\frac{1}{3}H_{sat}$.
Thus, a magnetization plateau should be observed over a relatively
wide field range.

\begin{figure}
\includegraphics[width=\columnwidth]{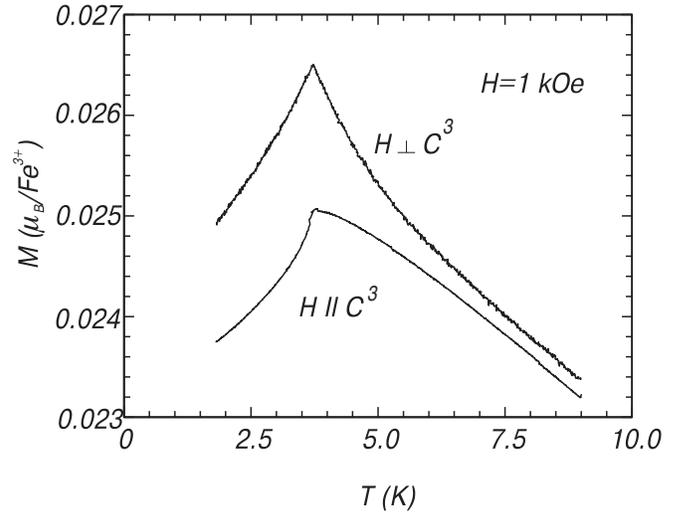}
\caption{ Temperature dependence of the magnetic susceptibility of
\RbFe\ for two directions of the magnetic field.}
\label{fig:susceptibility}
\end{figure}

In the case of an easy-plane magnetic anisotropy, an analogous
evolution of the sublattice orientations is expected when a
magnetic field is applied in the easy plane.  The umbrella-like
structure should be realized for magnetic fields applied along the
six-fold axis C$^6$, provided the easy-plane anisotropy is
sufficiently strong.

The magnetic resonance spectrum of a 120$^\circ$-spin structure
differs from the spectrum of a usual antiferromagnet both in the
number of normal modes and in the dependencies of frequencies on
the applied magnetic field. There are three eigenmodes for a 2D
triangular structure. \cite{AndreevMarchenko,Chubukov} For the
case of an easy-plane anisotropy, two of them are degenerate in
zero field, but have nonzero frequency; the third mode has
zero-frequency in zero field. In a magnetic field there are three
different resonance frequencies of uniform spin precession.

The presence of the interplanar antiferromagnetic exchange
requires one to consider at least a six-sublattice 3D magnetic
structures. For a weak interplanar exchange however, the main
features of the 2D-system should remain unchanged. The interplane
exchange may cause additional field-induced phase transitions,
where the mutual orientation of spins in neighboring planes
changes. Possible magnetic structures of the 3D $XY$ AFMT in a
magnetic field were analyzed in Ref.~\onlinecite{PlumerCalie},
while the Heisenberg AFMT was considered in
Ref.~\onlinecite{Gekht}. A two-fold period along the $C^6$ (or
$C^3$) axis is assumed here. The structures under consideration
are shown schematically in Fig.~1 and denoted as B1, B2, B3, C, D1
and D2. For the zero-field starting structure, B0, the nearest
neighboring spins, which are placed one above another along the
C$^6$-axis, are considered to be antiparallel due to the
antiferromagnetic interplane exchange. Depending on the relative
values of the in-plane magnetic field, interplane  and intraplane
exchange integrals, a phase from the set shown on Fig.~1 should be
realized. Thus, for a particular combination of both exchange
fields, one might expect a sequence of phase transitions in an
increasing magnetic field. Particularly, a sequence of transitions
like B1-B3-C-D1-D2-f is possible according to
Ref.~\onlinecite{Gekht}. The last phase corresponds to the
parallel orientation of the magnetization of all six sublattices
in a high magnetic field. Naturally, the interplanar interaction
may also modify the values of the critical fields $H_{c1}$ and
$H_{c2}$ with respect to the purely 2D case. \cite{Gekht}

As far as the magnetic resonance spectrum is concerned, the interplanar
exchange should result in a splitting of the main eigenfrequencies and in
the appearance of new modes due to the
increased number of sublattices.

There are several materials with triangular lattices carrying
magnetic ions, however, the members of the family AFe(TO$_4$)$_2$
(A=Cs, Rb; T=S, Mo) with \RbFe\ among them, are the most likely
candidates for AFMT-systems suitable for experiments. Other
related compounds,  e.g. the ABX$_3$-family (A=Cs, Rb; B=Ni, Mn,
Cu; X=Cl, Br, I), have an interplanar exchange that is larger than
the intraplanar one, thus they are quasi-1D magnets (see, e.g.,
Ref.~\onlinecite{Prozorova,Luthi}). ``Triangular" antiferromagnets
from other families, VX$_2$ (X=Cl, Br, I) (Ref.~\onlinecite{VX2})
and ACrO$_2$ with A=Li, Cu (Ref.~\onlinecite{ACrO2}), have
quasi-2D exchange, but their exchange field is too large, shifting
the field-induced transitions outside the range of conventional
measurements. A review of the magnetic properties of the
triangular magnets is given in Ref.~\onlinecite{Petrenko}.

In the present paper, we describe a study of the magnetic and
resonant properties of \RbFe, a material which can be considered
as a rare example of a nearly two-dimensional Heisenberg
antiferromagnet on a triangular lattice and that can be prepared
in single-crystal form. At room temperature the crystal structure
of \RbFe\ has the space group $P\bar{3}m1$. The magnetic Fe$^{3+}$
ions with spin $S=5/2$ are placed on the hexagonal lattice with
lattice parameters $a=$5.69~\AA\ and $c=$7.48~\AA. The
MoO$_4$-tetrahedra are placed between the layers of Fe$^{3+}$-ions
and form the structure with the three-fold axis. The exchange
integral $J$ representing the interaction within the planes should
be much larger than the exchange integral, $J^\prime$, of the
nearest neighbor ions in adjacent planes. The large difference in
these exchange integrals is due to the different exchange paths of
the indirect exchange interactions: via two oxygen ions within the
planes and via three or even more oxygen ions between the planes.
Thus, the structure of \RbFe\ may be considered (see, e.g.,
Ref.~\onlinecite{Inami}) as an ensemble of layers with a
triangular lattice occupied by $S=5/2$ Fe$^{3+}$ ions, and these
magnetic layers are separated by layers of MoO$_4$-Rb-MoO$_4$.
Magnetic ions in neighboring layers are placed one above another.

\begin{figure}[btp]
\includegraphics[width=\columnwidth]{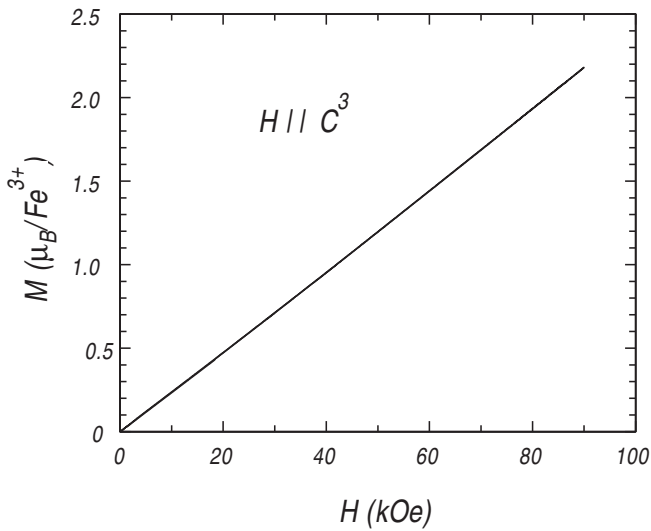}
\caption{\label{fig:Fig3} Field dependence of magnetization of
\RbFe\ for $H \parallel C^3$, $T=1.6$~K.}
\end{figure}

Evidence for a structural transformation at T=180~K was recently
reported in Ref.~\onlinecite{Popova}. Small changes of lattice
constants, Raman spectra and ESR linewidth indicate a structural
transformation, identified as a second-order phase transition from
a highly symmetric $P\bar{3}m1$ room-temperature structure into a
very similar but less symmetric $P\bar{3}c1$ low-temperature
structure. This transformation corresponds to rotations of the
MoO$_4$ tetrahedra. The "triangular" spin structure of
Fe$^{3+}$-layers in crystals of \RbFe\ was recently confirmed by
the elastic neutron scattering experiment in zero magnetic
field.\cite{Broholm} At the same time, according to the observed
neutron diffraction, an incommensurate modulation of the ordered
spin structure in the C$^3$-direction is present there: the mutual
orientation of spins from neighboring planes is close to the
antiparallel one but is slightly tilted at an angle of 17$^\circ$.
The low-temperature magnetization curves of powder samples of
\RbFe\ were reported earlier. \cite{Inami} The magnetization
saturated at a field of $H_{sat}=$~186 kOe and a magnetization
plateau marking the collinear phase was observed.

Neutron scattering experiments of Ref.~\onlinecite{Serrano}
confirming the triangular magnetic structure were performed for
powder samples of the related compounds CsFe(SO$_4$)$_2$ and
RbFe(SO$_4$)$_2$.

We have verified experimentally the theoretical concepts outlined
above by taking advantage of single-crystal samples of \RbFe. The
choice of a molybdate instead of a sulphate allowed us to avoid
the hydration and to obtain single crystals. Field-induced phase
transitions and low frequency spin dynamics in the different
phases were studied by means of magnetization measurements and ESR
spectroscopy.   In the
present paper we describe several field-induced phase transitions
including transitions not detected earlier. Each phase is found to
possess a characteristic set of spin-resonance modes.

\begin{figure}[btp]
\includegraphics[width=\columnwidth]{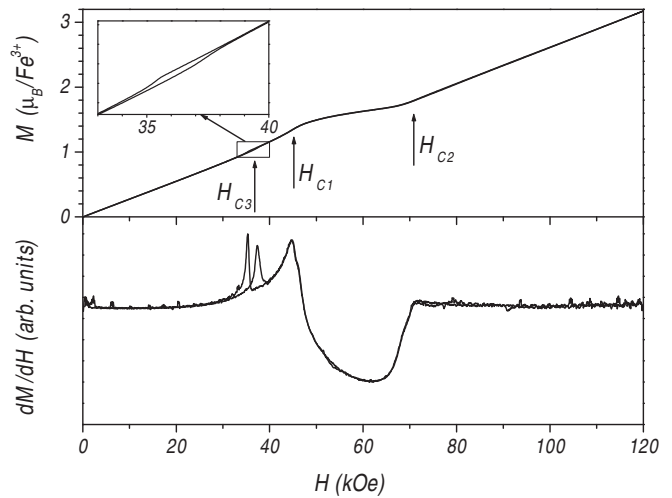}
\caption{\label{fig:Fig4} Field dependence of the magnetization
(top panel) and its derivative (bottom panel) of \RbFe\ for $H
\perp C^3$, $T=1.6$~K. The inset shows the hysteresis region
around the transition field $H_{c3}$.}
\end{figure}

\section{Samples and experimental techniques}

Single-crystal samples of \RbFe\ were synthesized by means of the
spontaneous crystallization from a flux melt. The mixture of
\RbFe\ and of K$_2$Mo$_2$O$_7$ in the molar ratio of 1:2 was
heated in a platinum crucible up to a temperature of 1300~K and
held at this temperature for 12 hours then cooled to 1000~K at a
rate of 3~K/h.  The nucleation of crystals was localized due to
the temperature gradient near a platinum rod which was inserted
into the melt in the precristallization state. The platinum rod
was withdrawn from the solution after the crystallization. The
K$_2$Mo$_2$O$_7$ flux was removed by dissolving in water. A much
slower dilution of the crystals of \RbFe\ takes place at the same
time. The crystals have the shape of thin hexagonal plates with
the size of 3-4 mm along each edge. The lattice parameters are in
accordance with those reported for powder samples. \cite{Inami}
The magnetization curves and the temperature dependencies of the
magnetic susceptibility were measured using a vibrating sample
magnetometer with the field range 0~-~120 kOe. Magnetic resonance
spectra were taken by a set of transmission-type magnetic
resonance spectrometers with resonators covering the range 9-120
GHz.

\section{Experimental results}

\subsection{Susceptibility and magnetization curves}

\begin{figure}[btp]
\includegraphics[width=\columnwidth]{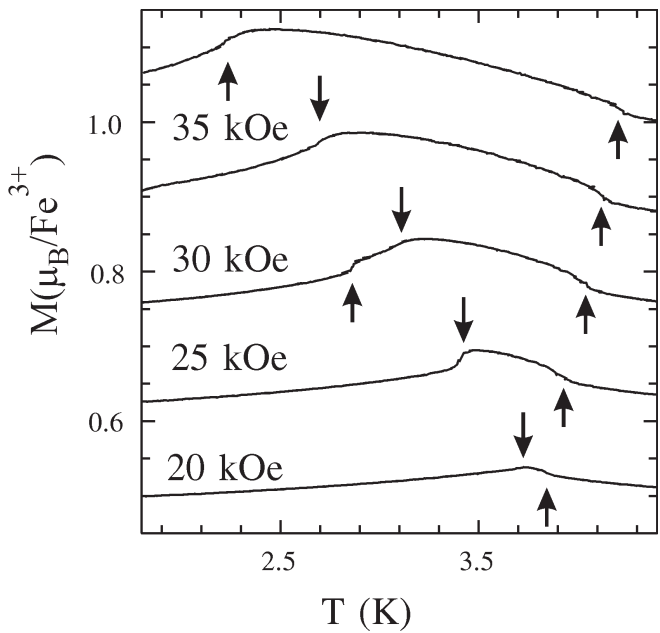}
\caption{\label{fig:Fig5} Temperature dependencies of the
magnetization of \RbFe\ in fixed magnetic fields. The arrows mark
the temperatures of the abrupt changes of the magnetization
corresponding to the magnetic phase boundaries.}
\end{figure}

The temperature dependence of the susceptibility of \RbFe\ at low
temperatures is shown in Fig.~2 and clearly demonstrates the
transition into the magnetically ordered state at $T_N=3.8$~K. The
susceptibility measurements in the whole range 10-300~K reveal the
temperature dependence of the Curie-Weiss type $(T+\Theta_c)^{-1}$
with the value of the Weiss constant $\Theta_c=22\pm2$~K. A small
(about 0.1\%) step-like anomaly in the reverse susceptibility
appears at 180~K giving an additional indication for the
structural transformation mentioned in the Introduction. There is
a significant deviation from the Curie-Weiss behavior below 10~K.
A strong anisotropy of the susceptibility also appears in this
temperature range, well above $T_N$.

Below the transition point $T_N=3.8$~K, the magnetization curves
$M(H)$ are quite different for different orientations of the
magnetic field. For $H \parallel C^3$,  the magnetization
increases linearly with the field for all values of  the applied
field (see Fig.~3). For $H\perp C^3$, the field dependence of the
magnetization is much more complicated, as shown in Fig.~4 for
$T=1.6$~K. There are abrupt changes in the slope of the
magnetization curve at 47~kOe and 71~kOe, with the differential
magnetic susceptibility being significantly reduced in the region
between these fields.

An additional, first-order phase transformation with a hysteresis
in $M(H)$ curve was detected at $H_{c3}=35$ kOe. Magnetization
measured between 30 and 38 kOe shows a large difference in signal
obtained for the rising and falling magnetic field. A smaller, but
still clearly observable difference exists down to a field of
20~kOe. In order to detect the transitions by changing temperature
we measured the temperature dependence of the magnetic moment at
constant magnetic field. Several examples of these data are shown
in Fig.~5. The temperatures of the abrupt changes of the
magnetization curves are marked by arrows and indicate the field
dependence of the N\'{e}el temperature and the temperature
dependence of the fields $H_{c1}$ and $H_{c3}$. The values of
critical magnetic fields and temperatures derived from the curves
like those shown in Figs.~4 and 5 are collected on the $H-T$ phase
diagram in Fig.~6. There are  at least 4 ordered antiferromagnetic
phases P1, P2, P3, P4 and a paramagnetic phase PM.

It should be noted that smeared changes in the magnetization slope
were observed at the critical fields $H_{c1}$ and $H_{c2}$ for
temperatures slightly above the N\'{e}el temperature. These
regions are marked in Fig.~6 by shadowed ovals.

\begin{figure}[btp]
\includegraphics[width=\columnwidth]{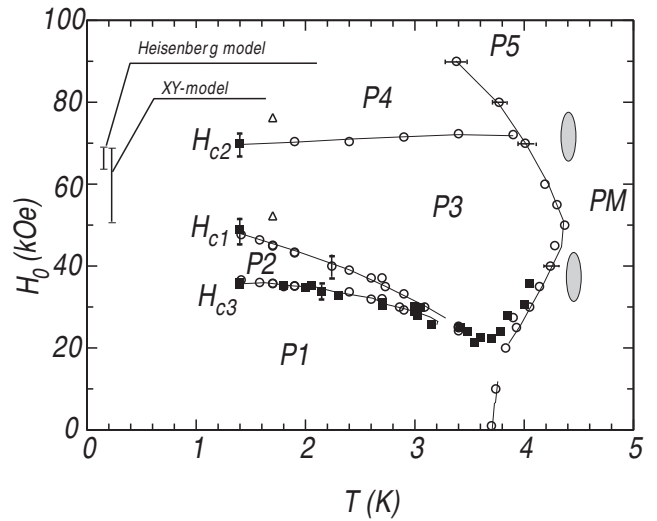}
\caption{\label{fig:Fig6} The $H-T$ magnetic phase diagram of
\RbFe\ derived from the single crystal magnetization and ESR
measurements. Data marked by triangles are taken from powder
measurements \cite{Inami}, open circles correspond to
magnetization measurements, filled squares - to ESR-data. The two
vertical line segments near $T=0$ illustrate the field ranges of
the phase ``c" according to calculations from
Ref.~\onlinecite{Chubukov} for the Heisenberg and the XY-model.
The two shadowed ovals mark the regions of smeared changes in the
magnetization slope for $T>T_N$.}
\end{figure}

\subsection{Antiferromagnetic resonance}

On cooling the samples, the ESR line broadens markedly below 10~K.
After passing through the N\'{e}el point, the ESR line shifts from
the paramagnetic resonance position and again becomes rather
narrow. The temperature evolution of the ESR line is shown in
Fig.~7. In this paper we discuss only the low-temperature magnetic
resonance, well below the N\'{e}el point. The critical behavior
will be the subject of further investigations.

The field-dependencies of the microwave transmission at $T=$~1.3~K
are shown in Fig.~8 and Fig.~9 for two orientations of the
magnetic field. We can derive the field dependencies of the
spin-resonance modes in the ordered state from the positions of
the resonance-absorption lines at different frequencies. As shown
in Fig.~9, the microwave absorption data are sensitive to all the
phase transitions detected in the magnetization measurements. The
hysteresis loop around the field $H_{c3}$ is clearly evident in
the microwave absorption. The change of the microwave absorption
at this transition is frequency-dependent. The curves taken at a
frequency around 25 GHz are the most sensitive to this transition
as they demonstrate the largest hysteresis around  $H_{c3}$.

\begin{figure}
\includegraphics[width=\columnwidth]{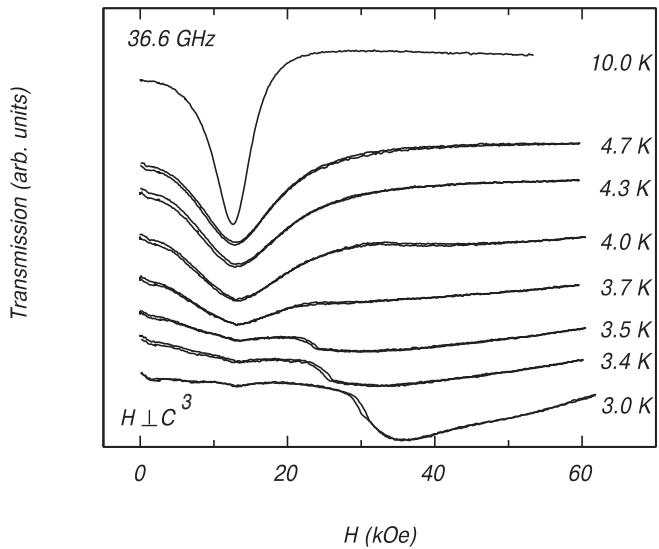}
\caption{\label{fig:Fig7} Temperature evolution of the 36.6 GHz
ESR line in \RbFe\ at cooling through the critical region and
N\'{e}el point.}
\end{figure}

For $H\parallel C^3$, two branches of the resonance are clearly
seen. The frequency of the first branch rises with the field,
while the frequency of the second branch decreases (see Figs.~8
and 10). The frequencies of these two branches are monotonic
functions of the applied magnetic field as expected for the
umbrella-like structure.

For $H\perp C^3$, a complicated nonmonotonic dependence of the
resonant frequencies $\nu_i$  with varying magnetic field was
observed (see Fig.~11). The frequency-field curves contain abrupt
changes at the fields of the phase transitions in accordance with
the data of the magnetization measurements. The values of the
phase-transition fields derived from the microwave-absorption
curves are marked on the phase diagram in Fig.~6 by filled squares
and are in a good agreement with the results of static
magnetization measurements. The total number of the observed spin
resonance modes is five.

\section {Discussion}

\subsection{Basic principles}

The observation of the sequence of the phase transitions for the
magnetic field lying in the basal plane implies that there is an
easy-plane type of anisotropy (an easy-axis anisotropy would
result in the umbrella-like spin structure at $H\perp C^3$ without
the cascade of field-induced phase transitions).

\begin{figure}
\includegraphics[width=\columnwidth]{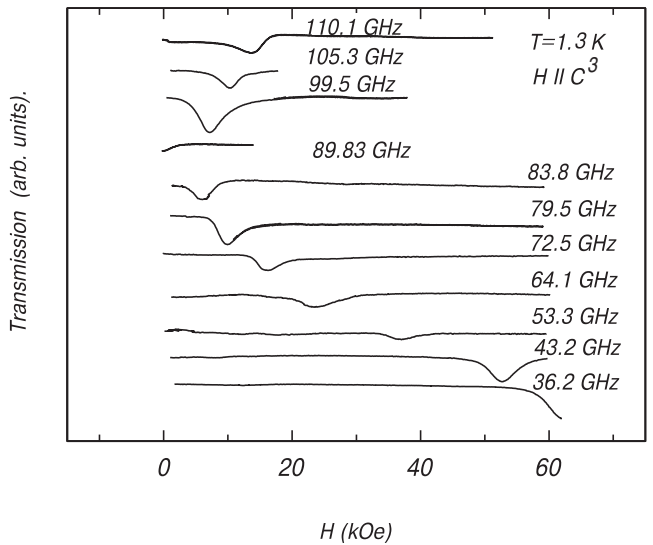}
\caption{\label{fig:Fig8} Magnetic resonance lines in \RbFe\ at
different frequencies at $H\parallel C^3$, $T=1.3$~K.}
\end{figure}

At the moment there is no explanation for the nature of the
incommensurate modulation observed in Ref.~\onlinecite{Broholm},
but on the assumption of the small deviation from the antiparallel
orientation of spins in neighboring planes, we shall describe the
field-dependent spin structure in the approximation of the perfect
antiferromagnetic orientation of spins in neighboring layers in
the zero-field ground state.

Thus, we consider the following model spin-Hamiltonian, following
the notation of Ref.~\onlinecite{Chubukov}.
\begin{eqnarray}
{\cal
H}=2J\sum_{(ij),n}S_{in}S_{jn}+2J'\sum_{in}S_{in}S_{in+1}\nonumber
\\
+D\sum_{in}{(S^z_{in})^2} -g\mu_BH\sum_{in}S_{in}.
\end{eqnarray}
Here the sums are taken within the layers ($i,j$) and along the
transverse direction ($n$), and $D>0$ is the constant of the
anisotropy of the easy-plane type.

\begin{figure}
\includegraphics[width=\columnwidth]{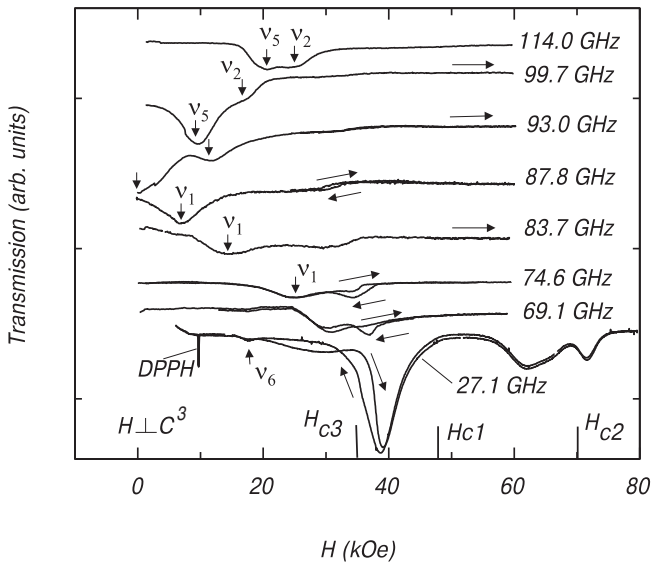}
\caption{\label{fig:Fig9} Magnetic resonance lines in \RbFe\ at
different frequencies at $H\perp C^3$, $T=1.3$~K.}
\end{figure}

\subsection{Field-induced phase transitions and molecular fields}

Using the molecular-field calculations \cite{Chubukov} of the
susceptibilities along and perpendicular to the $C^3$-axis
\begin{eqnarray}
\chi_{\parallel} & = & \frac{{(g\mu_B)}^2}{18J+2D}N, \nonumber \\
\chi_{\perp} & = & \frac{{(g\mu_B)}^2}{18J}N, \nonumber
\end{eqnarray}
($N$ is the number of magnetic ions), we can estimate the strength
of the exchange field, $H_E$, and the anisotropy field, $H_A$:
\begin{eqnarray}
H_E & = & \frac{6JS}{g\mu_B} = 67 {\rm ~kOe}, \nonumber \\
H_A & = & \frac{DS}{g\mu_B} = (5\pm 1.5) {\rm ~kOe}. \nonumber
\end{eqnarray}
The value of $H_E$ is in a good agreement with the saturation
field value $H_{sat}$=186 kOe measured in Ref.~\onlinecite{Inami}
(according to the molecular-field calculations $H_{sat}=3H_E$).

The nearly vanishing slope in the $M(H)$-curve in the field range
between $H_{c1}$ and $H_{c2}$ indicates the presence of
magnetization plateau and suggests that we should treat this phase
as the collinear phase (phase C of Fig.~1), which is stabilized in
the range $H_{c1}<H<H_{c2}$ by thermal and quantum fluctuations.
\cite{Chubukov,Korshunov,Rastelli12} Using the $J$ value derived
from the susceptibility measurements and the ratio $J^\prime/J$
derived from the magnetic resonance spectra (see the next section)
we can propose a sequence of phase transition on the basis of the
model of Ref.~\onlinecite{Gekht}, where the calculated phase
diagram at $T=0$  is plotted on the $H/J, J^\prime S/J$ - plane.
At the consideration of that phase diagram one should remember
that our situation of the easy-plane anisotropy should be
considered neglecting the umbrella-like phase derived for the
isotropic approximation. For our case of $J^\prime S/J=0.095$ with
an easy plane anisotropy this phase diagram predicts a sequence of
phase transitions B1-B2-C-D1-D2 in the magnetic field. Thus,
according to the theoretical analysis of Ref.~\onlinecite{Gekht},
we can propose that in the field $H_{c3}$ we have transition like
B1-B2 with change in the mutual orientation of spins in
neighboring layers, then transition B2-C at $H=H_{c1}$ and C-D1 at
$H=H_{c2}$. Thus we propose the observed sequence P1-P2-P3-P4 may
be treated as B1-B2-C-D1.

It should be noticed that the phase boundary observed in the
temperature range 3.4-4.2~K and in fields above $H_{c2}$ cannot be
smoothly extrapolated to the saturation field 186~kOe at
$T$=1.3~K. Therefore, the existence of yet another phase
transition at higher magnetic field cannot be ruled out. According
to the analysis given in Ref.~\onlinecite{Gekht}, the phase D2 is
energetically favourable for certain values of the interplanar
exchange in a field just below the saturation field. Thus we
suggest the region of the phase diagram indicated in Fig.~6 as P5
may be treated as the D2 phase.

Using only the results of magnetization and ESR measurements we
cannot distinguish between the  phases B1, B2, B3 and between the
D1 and D2 phases. Thus, an alternative sequence of phase
transitions, such as B2-B3-C-D1 (Ref.~\onlinecite{Korshunov2})
should, in principle, also be considered as an explanation of the
observed sequence P1-P2-P3-P4. A scenario for the field-induced
transitions with B2 as the starting phase is also suggested on the
basis of the molecular field approximation in
Ref.~\onlinecite{PlumerCalie}. The tiny differences between the
free energies of the phases B1 and B2 may be associated with the
contribution of thermal and quantum fluctuations which should be
taken into account along with the anisotropy. In any case, the
exact solution of this problem will only have a limited
significance for \RbFe\ because of the incommensurate modulation
mentioned, which is still not included in the theoretical models.

Nevertheless, the observation of the $H_{c3}$-phase transition
clearly marks the effect of the interplanar exchange on the
field-dependent phases of the AFMT with weakly coupled layers.

The observed boundaries in the collinear phase range may be
compared to the calculations \cite{Chubukov} performed using an
isotropic ($D=0$) and XY (infinite $D$) models at $T=0$~K. The
calculated zero-temperature field-ranges of the C-phase (using
$H_E=67$~kOe) are shown in the phase diagram in Fig.~6 by the line
segments. There is a qualitative agreement with our observations.

\begin{figure}
\includegraphics[width=\columnwidth]{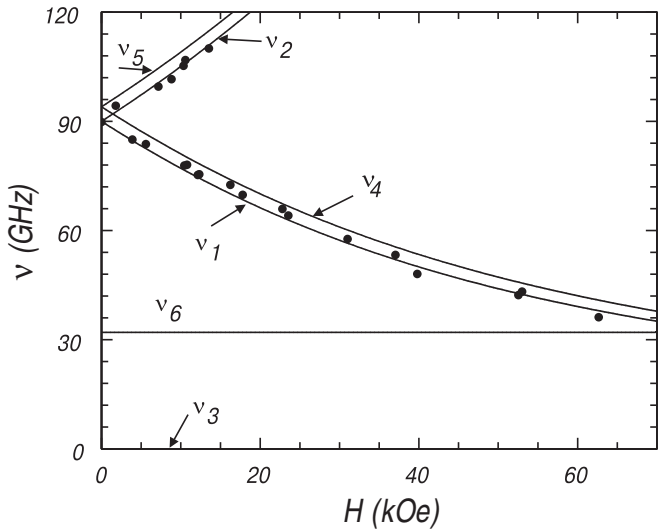}
\caption{\label{fig:Fig10} ESR frequency of \RbFe\ vs. magnetic
field for $H \parallel C^3$, $T=1.3$~K. The curves are calculated
dependencies after (11, 12),  parameters are described in the
text.}
\end{figure}

\subsection{Spin-resonance modes}

For a triangular system with antiferromagnetic interplane exchange
interactions, there should be five resonance modes with non-zero
frequencies (the frequency of the sixth mode is zero in small
fields in absence of any in-plane anisotropy). We have observed
all five resonances for $H \perp C^3$. For weak magnetic fields,
where the exchange triangular spin configuration is only slightly
distorted, the resonance frequencies may be calculated following
the macroscopic theory based on a classical Lagrange formalism
\cite{AndreevMarchenko}. This method of calculation is suitable
for complicated and multi-sublattice systems, however it is valid
only in the field range where the exchange magnetic structure is
not strongly distorted by the magnetic field.~\footnote{Authors
acknowledge V.~I.~Marchenko for making these calculations.} The
basic principles of the calculations and the resulting formulae
are given in Appendix A, while the field-dependencies of the
resonant frequencies $\nu_{i}$ are presented in Fig.~10, and
Fig.~11 (the values of the parameters will be described below).
For the modes $\nu_{3,6}$ in the low-field range, the oscillating
components lie within the plane. The modes with frequencies
$\nu_{1,2,3}$ correspond to the in-phase oscillations of spins in
the neighboring planes while for the modes with frequencies
$\nu_{4,5,6}$, the spins in the neighboring planes oscillate out
of phase. Note that for a purely 2D AFMT only the modes
$\nu_{1,2,3}$ are present.

\begin{figure}
\includegraphics[width=\columnwidth]{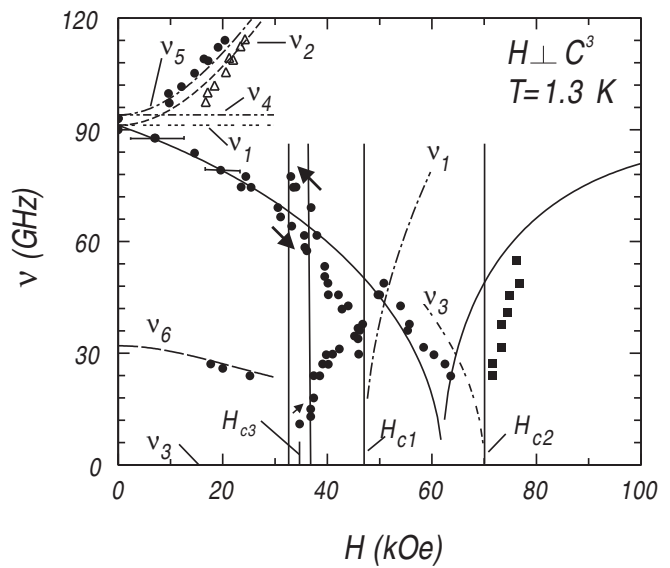}
\caption{\label{fig:Fig11} ESR frequency of \RbFe\ vs. magnetic
field for $H \perp C^3$, $T=1.3$~K. Dashed curves represent the
calculation after (8, 9, 19, 20), solid curves -- after (13, 18).}
\end{figure}

The frequencies calculated in this way are only valid for small
magnetic fields, when the exchange triangular structure is not
strongly distorted. Thus the approximation used is not strictly
true in fields of the order of $H_{c1}$. The distortion of the
triangular structure, as shown in the 2D approximation
\cite{Chubukov}, will result in a field dependence of the
resonance frequencies $\nu_1$, $\nu_3$. The approximate
correspondence between the parameters of the macroscopic theory
described in Appendix A and of the spin-Hamiltonian may be derived
using estimates (i.e. neglecting zero-point fluctuations) for the
susceptibility and resonance frequencies derived for 2D AFMT in
the molecular-field approximation:
\begin{eqnarray}
\label{a_square}
 a^2 & = & 18JDS^2= \gamma^2 3 H_AH_E,\\
 c^2 & = & \gamma^2 H_E^2 \frac{J^\prime}{J},
\label{b_square}
\end{eqnarray}
here $\gamma = g \mu_B/2 \pi \hbar$.

The nonzero frequency of the mode $\nu_6$ confirms the interplanar
exchange. Another indication of the interplanar exchange is the
splitting of the rising ESR mode (the difference in resonance
fields of the modes $\nu_2$ and $\nu_5$ in Fig.~11).

The five spin resonance branches shown in the low-field range in
Fig.~11 and the resonance frequencies  for ${\bf H} \parallel C^3$
(Fig.~10) may be reasonably described by the equations (8-12) with
only two fitting parameters:  $a$=90.8 GHz and $c$=32 GHz. The
dashed curves in Fig.~11 and curves in Fig.~10 are calculated in
this way with the anisotropy parameter of the susceptibility,
$\eta=0.05$, derived from the susceptibility measurements.

Using the value of the susceptibility described above and the
molecular-field relations (\ref{a_square}, \ref{b_square}) we can
evaluate the molecular fields and parameters of the Hamiltonian:
$H_E= 67$ kOe, $H_A= 5.2$ kOe, $J^\prime/J=0.039$. These values of
$H_E$ and $H_A$ are in reasonable agreement with the values
derived from the susceptibility and saturation field.

For $H\perp C^3$, the triangular spin structure of \RbFe\ is
already strongly distorted in a field of about 10~kOe. Therefore
the macroscopic theory of magnetic dynamics given in
Ref.~\onlinecite{AndreevMarchenko} cannot be used. To have at
least an approximate description of the resonant frequencies over
wider range of applied fields, we used the calculations for the
three-sublattice 2D model  (i.e. for $J^\prime=$0) in a classical
approximation. \cite{Chubukov} The results of these calculations
(relations 13, 18) are presented in Fig.~11 by the solid lines.
The values of $D$ and $J$ used to calculate these curves are taken
from the values of $H_E$ and $H_A$ given above and thus agree with
the parameters used for fitting the data in low fields according
to relations (8-12).

For qualitative description of the resonance frequencies within
the field range of the collinear phase $H_{c1}<H<H_{c2}$, one
should take into account the zero-point fluctuations stabilizing
this phase in that field range. The appropriate calculation was
made in the $J^\prime=0$ approximation with the assumptions of the
XY-model in the same paper of Chubukov and Golosov.
\cite{Chubukov} Corresponding relations  for the lowest resonance
frequencies are given in  Appendix B (relations 19,20) and are
plotted by dashed lines in the region $H_{c1}<H<H_{c2}$ in
Fig.~11. These formulae reveal critical softening of the resonance
frequencies near the fields $H_{c1}$ and $H_{c2}$. We observe a
soft mode near $H_{c2}$ which is in qualitative agreement with
relation (\ref{XYmodel}).

\section{Conclusion}

A sequence of  field-induced phase transitions, governed by the
intraplane exchange interaction and by the weak interplane
exchange was found in \RbFe, a quasi-2D antiferromagnet on a
triangular lattice. The magnetic properties may, in part, be
explained in terms of a 2D antiferromagnet on a triangular
lattice. However, a phase transition governed by weak interplane
exchange was found. This phase transition is accompanied by the
changes in spin-resonance spectra. A self-consistent description
of the magnetization curves, phase transitions and resonance modes
in a wide field range is given.

\begin{acknowledgments}
We kindly acknowledge V.~I.~Marchenko for making calculations of
resonance spectra and discussions, S.~E.~Korshunov,
M.~E.~Zhitomirsky,  A.~V.~Chubukov and S.~S.~Sosin for numerous
discussions, M.~R.~Lees for critical reading of the manuscript and
C.~Broholm for transmitting to us the results of the unpublished
work. \cite{Broholm} We acknoledge the support by the Russian
Foundation for Basic Research (RFBR), grant No.01-02-17557, and by
the Award No. RP1-2097 of the U.S. Civilian Research and
Development Foundation for the Independent States of the former
Soviet union (CRDF).
\end{acknowledgments}

\section{Appendix A. Spin resonanse modes in the macroscopic approach}

We consider 120$^\circ$-triangular structures in planes 1 and 2,
which can be represented by normalized spin densities
\begin{equation}
\begin{array}{c}
{\bf S}_1({\bf r}) = {\bf l}_{11} \cos {\bf qr} +{\bf l}_{12} \sin
{\bf qr}, \\ {\bf S}_2({\bf r}) = {\bf l}_{21} \cos {\bf qr} +{\bf
l}_{22} \sin {\bf qr}.
\end{array}
\end{equation}
Here ${\bf l}_{11}, {\bf l}_{12}$ and ${\bf l}_{21}, {\bf l}_{22}$
are two pairs of  orthogonal unit vectors of antiferromagnetism,
$q_x=4\pi/3a$, $q_y=0$, $a$ - crystal period within the plane;
vector ${\bf r}$ takes discrete values within the basal plane
pointing at the magnetic sites on a triangular lattice. This
structure is due to the in-plane exchange. The mutual orientation
of spin triangles in neighboring planes for weak interplane
exchange corresponds to the minimum of the Heisenberg exchange
energy of nearest neighbors from different planes:
\begin{equation}
\alpha ({\bf l}_{11}{\bf l}_{21}+{\bf l}_{12}{\bf l}_{22})
\label{interplane}
\end{equation}

For $\alpha>0$: ${\bf l}_{11}=-{\bf l}_{21}$, ${\bf l}_{12}=-{\bf
l}_{22}$.

For $\alpha<0$: ${\bf l}_{11}={\bf l}_{21}$,  ${\bf l}_{12}={\bf
l}_{22}$.

Note, that the low-frequency spectrum, which depends on the
quadratic expansion on the angles of the mutual rotation of the
spin triangles, is the same for the antiferromagnetic ($\alpha>0$)
and ferromagnetic ($\alpha<0$) exchange.

The Lagrange function of the spin dynamics has three
contributions: two ``ordinary" terms of the triangular structures
of the two systems of spin planes $({\bf l}_{11},{\bf l}_{12})$
\begin{equation}
\frac{1}{4}\left
\{\frac{\chi_\perp}{\gamma^2}(\mathbf{\Omega}_1+\gamma {\bf H})^2
+ \frac{\chi_\parallel
-\chi_\perp}{\gamma^2}(\mathbf{\Omega}_1+\gamma {\bf H},{\bf
n}_1)^2 +\beta n_{1z}^2  \right \}
\end{equation}
and $({\bf l}_{21},{\bf l}_{22})$
\begin{equation}
 \frac{1}{4} \left \{ \frac{\chi_\perp}{\gamma^2}(\mathbf{\Omega}_2+\gamma {\bf H})^2 +
 \frac{\chi_\parallel -\chi_\perp}{\gamma^2}(\mathbf{\Omega}_2+\gamma {\bf H},{\bf n}_2)^2 +\beta n_{2z}^2  \right \}
\end{equation}
while the third term, the energy of the interplane exchange, is
given above (\ref{interplane}). Here ${\bf n}_1=[{\bf l}_{11}{\bf
l}_{12}]$, ${\bf n}_2=[{\bf l}_{21}{\bf l}_{22}]$ are unit vectors
in the spin space, which are normal to the spin planes, $\beta$ --
is the constant of uniaxial anisotropy, $\mathbf{\Omega}_1$ and
$\mathbf{\Omega}_2$ -- are angular velocities of rotation of the spin
triangles. The values $\chi_\parallel$ and $\chi_\perp$- determine
the components of the tensor of the magnetic susceptibility in the
ground state, along and perpendicular to the vector ${\bf n}_1$ respectively.

For the in-plane field, the frequencies of the in-phase
oscillations of spins in neighboring planes are given by:
\begin{equation}
\begin{array}{l}
\nu_1=a, \\ \nu_2=\sqrt{a^2+H^2}, \\ \nu_3=0
\end{array}
\end{equation}

The additional frequencies due to the interplane exchange (spins
of different planes oscillate "out of phase") are
\begin{eqnarray}
\nu_4  =  b = \sqrt{a^2+A^2} & & \nonumber  \\ 2\nu^2_{5,6}  =
b^2+c^2+H^2 \pm & & \\ & \hspace*{-20 mm}
\sqrt{(b^2+c^2+H^2)^2-4c^2(b^2-\eta H^2)} &  \nonumber
\end{eqnarray}
Here $A^2=\gamma^2 \alpha/\chi_\perp$, $a^2 = \gamma^2
\beta/\chi_{\perp}$,
$\eta=\frac{\chi_\parallel-\chi_\perp}{\chi_\perp}$,
$c^2=2A^2/(1+\eta)$. Parameters $a$, $b$ and $c$ are coupled by
the relation
\begin{equation}
2(b^2-a^2)=(1+\eta)c^2
\end{equation}

For the magnetic field oriented parallel to the $C^3$-axis,
ordinary resonances (in-phase motion of triangles) are:
\begin{equation}
\begin{array}{l}
\nu_{1,2}=\sqrt{a^2+(\frac{1+\eta}{2}\gamma
H)^2}\pm\frac{1-\eta}{2}\gamma H, \\ \nu_3=0,
\end{array}
\end{equation}
while the additional frequencies (the triangles rotate in opposite
directions) are:
\begin{equation}
\begin{array}{l}
\nu_{4,5}=\sqrt{b^2+(\frac{1+\eta}{2}\gamma
H)^2}\pm\frac{1-\eta}{2}\gamma H, \\ \nu_6=c
\end{array}
\end{equation}

\section{Appendix B. Spin-resonance modes in 2D model \cite{Chubukov}}
\subsubsection{Molecular field approximation}

Introducing the normalized field
\[ h=\frac{2\mu_BH}{6JS}=3H/H_{sat} \]
we have the following resonance frequencies, for the magnetic
field lying in the easy plane of 2D-AFMT: \\ 1) below the
transition to the collinear phase ($0<h<1$)
\begin{eqnarray}
\nu_1 & = &
\frac{6JS}{2\pi\hbar}\left[\frac{D}{6J}(3-2h-h^2)\right]^{\frac{1}{2}}
\\ \nu_2 & = &
\frac{6JS}{2\pi\hbar}\left[\frac{D}{6J}(3+2h+h^2)\right]^{\frac{1}{2}}
\\ \nu_3 & = & 0
\end{eqnarray} \\
2) above the transition to the canted phase ($1<h<3$)
\begin{eqnarray}
\nu_1 & = & 0 \\ \nu_2 & = & \frac{6JS}{2 \pi
\hbar}\left[\frac{D}{6J}
\frac{(h^6-3h^4+35h^2+63)}{16h^2})\right]^{\frac{1}{2}} \\ \nu_3 &
= & \frac{6JS}{2 \pi \hbar}\left[\frac{D}{6J}
\frac{(9-h^2)(h^2-1)(h^2+7)}{16h^2})\right]^{\frac{1}{2}}
\end{eqnarray}

\subsubsection{XY-model including fluctuations}

The two lowest resonant frequencies in the collinear phase are
\begin{eqnarray}
\nu_1 & = &
\frac{6JS}{2\pi\hbar}\left(h-h_{c1}\right)^{\frac{1}{2}}, \\ \nu_3
& = &
\frac{6JS}{2\pi\hbar}\left(\frac{h_{c2}-h}{3}\right)^{\frac{1}{2}}.
\label{XYmodel}
\end{eqnarray}
Here  $h_{c1}$ and $h_{c2}$ are the normalised values of the
critical fields $H_{c1}$ and $H_{c2}$.

\end{document}